\documentclass[
aps,%
12pt,%
final,%
notitlepage,%
oneside,%
onecolumn,%
nobibnotes,%
nofootinbib,%
superscriptaddress,%
centertags]%
{revtex4}
\usepackage{layout}
\usepackage[latin1]{inputenc}
\usepackage[english]{babel}
\usepackage[T1]{fontenc}
\usepackage{amsmath, latexsym, soul, color, type1cm,dsfont, float, fancyhdr}
\usepackage[official]{eurosym}
\usepackage{graphicx}
\usepackage{amsmath}
\usepackage{amssymb}
\usepackage{graphicx}

\usepackage{amsmath}
\usepackage{amssymb}
\usepackage{graphicx}
\usepackage{sidenotes}
\usepackage{pstricks}
\usepackage{fancybox}
\usepackage{xspace}%
\usepackage[varg]{txfonts}
\begin{document}
%
\title{Long-range rapidity correlations in high energy AA collisions
 in Monte Carlo model with string fusion}
%
%

\author{\firstname{Vladimir}~\surname{Kovalenko}}
\email{nvkinf@rambler.ru}
\author{\firstname{Vladimir}~\surname{Vechernin}}
\affiliation{Faculty of Physics, St.Petersburg State University, Ulyanovskaya st. 3, 198504, St.Petersburg, Russia
}

\begin{abstract}{The magnitude 
of long-range correlations between observables in two separated rapidity windows, proposed as a signature of the string fusion and percolation phenomenon, is studied in the framework of non-Glauber Monte Carlo string-parton model, based on the
picture of elementary collisions of color dipoles.  The predictions, obtained with and without string fusion, demonstrate effects of color string fusion on the observables in Pb-Pb collisions at the LHC: decrease of n-n correlation coefficient with centrality and negative
   pt-n correlations, if the sufficiently effective centrality estimator is applied.
In general case 
it is shown that the values of n-n and pt-n correlation coefficients
strongly depend on the method of collision centrality fixation. 
In contrast, the predictions obtained for pt-pt correlation have almost no effect of centrality determination method and the corresponding experimental data would 
produce the strong limitation on the transverse radius of a string.
}
\end{abstract}

%
\maketitle
\section{Introduction}
\label{intro}
The study of long-range correlations between observables in two windows separated in
rapidity was proposed \cite{Amelin} as a signature of the string fusion and percolation phenomenon, which is
one of the collectivity effects \cite{BraunPajares} in ultrarelativistic heavy ion collisions. It was
suggested to consider different types of correlations \cite{ALICEtdr}: n-n, pt-n, pt-pt, where n is the event multiplicity of charged particles in a given rapidity window and pt is their event mean transverse momentum: $p_t=(1/n) \sum_{i=1}^n p_{t_i}.$
The correlation coefficient is defined as the slope of the correlation function in normalized variables \cite{VechKol0}:
${b_{BF}=\frac{ \langle F \rangle }{\langle B \rangle }
 \frac{{d}\langle B \rangle_{F}}{dF}|_{F=\langle F \rangle}}$, where $B$ and $F$ could be multiplicity or $p_t$. Multiplicity fluctuations are defined as scaled
 variance of the charged multiplicity $w={\text{Var } n / \langle n \rangle}$.

The values of correlation coefficients in AA collisions could strongly depend 
on the method of collision centrality fixation \cite{VechKol0}. 
The present work is devoted to study of the correlation coefficients in Pb-Pb
collisions at the LHC energy in the framework of the non-Glauber Monte Carlo string-parton model,
and obtaining the predictions for the LHC in the conditions close to experimental.

\section{Non-Glauber Monte-Carlo model 
\label{sec-1}
}
The present model \cite{KovalenkoYaF,KovalenkoPoS} is based on partonic picture of nucleons interaction,
accounting the energy and angular momentum conservation in the initial state of a nucleon.
The nucleon is assumed to be composed of several set of pairs (quark-diquark and quark-antiquark), forming dipoles. The probability of dipoles interaction depends on the transverse coordinates of their ends \cite{Lonnbland,Gustafson} with some effective coupling constant.
Multiplicity and transverse momentum are calculated in the approach of colour strings, stretched between projectile and target partons, taking into account their finite rapidity width and interaction in transverse plane -- string fusion \cite{BraunPajaresVechernin,BraunAll}.
The main effect of the fusion of strings is modification of the mean transverse momentum and multiplicity, coming from
a cluster of overlapping strings:
\begin{equation}\nonumber
\langle \mu \rangle_k = \mu_0 \sqrt{k} \frac{S_k}{\sigma_0}, \hspace{1cm} \langle p_t^2 \rangle_k = p_0^2 \sqrt{k},  \hspace{1cm} 
\langle p_t \rangle_k = p_0 \sqrt[4]{k},
\end{equation}
where $S_k$ --
 area, where k strings are overlapping, $\sigma_0$ -- single string transverse area, $\mu_0$ and $p_0$ -- mean multiplicity and transverse momentum from one string.
A string (or cluster of overlapping strings) fragments independently in any rapidity interval according to Poisson distribution.
The model parameters are constrained from the p-p data on total inelastic cross section and multiplicity
and multiplicity in minimum bias p-Pb and Pb-Pb collisions \cite{KovalenkoPoSMC}.

For the centrality estimation we used two approaches: number of participant nucleons ($N_{\text{part}}$)
and multiplicity estimator (``vzero''), covering multiplicity in rapidity windows:
 (3.0; 5.0) and (-3.6; -1.6), that is close to acceptance
of ALICE detector V0 \cite{ALICEV0}, used for the centrality determination \cite{ALICECent}.

\section{Results}
\label{sec-2}
\begin{figure}[h]
\centering
\includegraphics[width=3.64cm,clip]{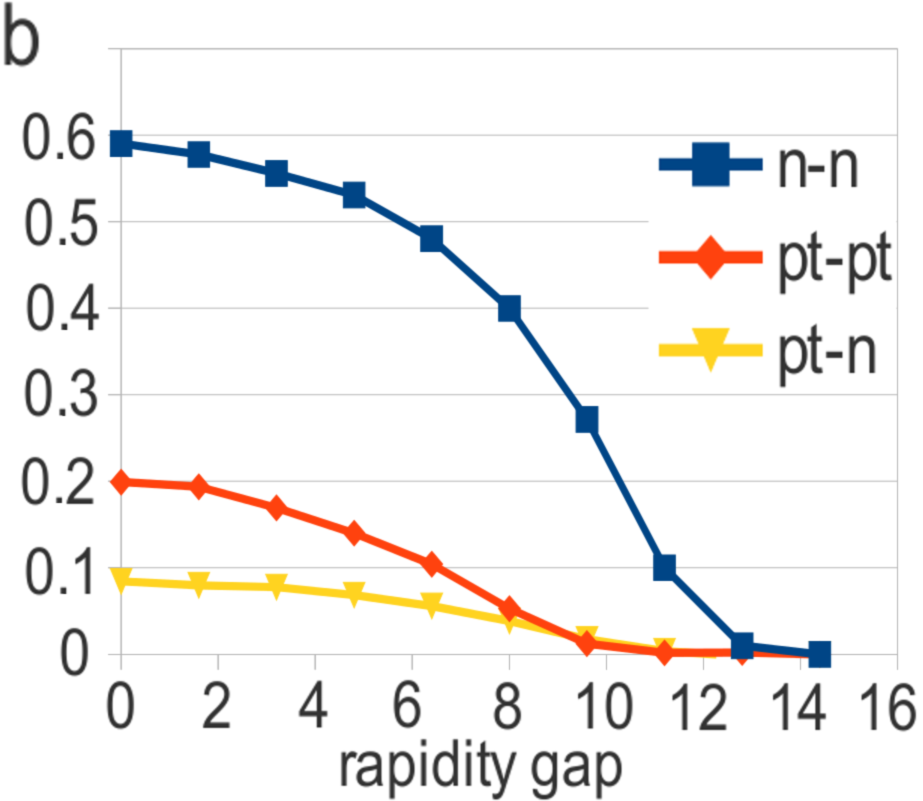} \ 
\includegraphics[width=7.56cm,clip]{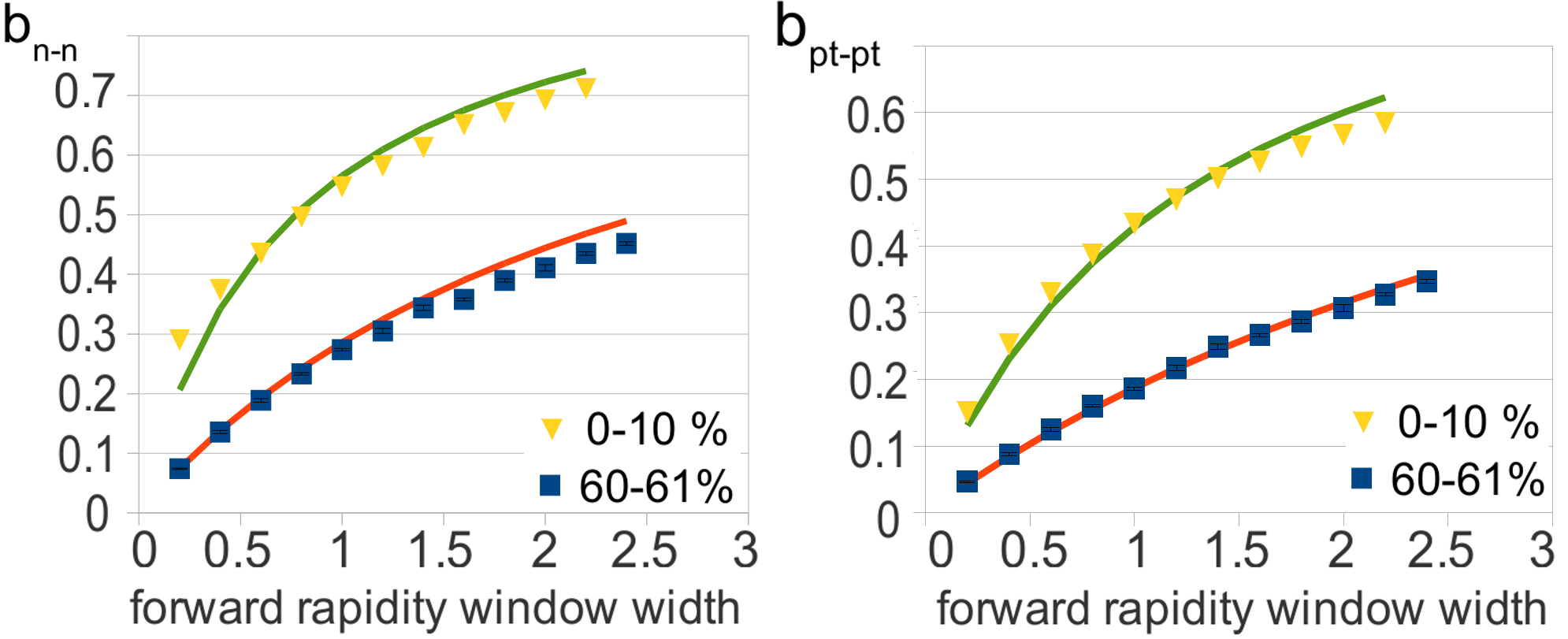}
\caption{Dependence of the correlation coefficients on the rapidity gap between windows (left) and on the width of
the forward window of $b_{n\text{-}n}$ (middle) and $b_{pt\text{-}pt}$ (right). }
\label{fig-1}       
\end{figure}

The correlation coefficients in Pb-Pb collisions at 2.76 TeV in two rapidity windows width of 0.8 as a function of
the gap between them are shown in fig. \ref{fig-1} (left).

We can observe almost flat behaviour in mid-rapidity and decrease of correlations with further increase of the gap.
Both pt-n and pt-pt correlations vanish at gap about 10 rapidity units, while n-n correlations are still present, 
and this behaviour is similar to pp case \cite{KovalenkoPoS}.

The dependence of the correlation coefficients
on the forward rapidity windows width is shown in the middle and right part of fig. \ref{fig-1}.
 The results demonstrate noticeable dependence of $b_{n\text{-}n}$ on the forward rapidity window size ${\Delta y_F}$.
The points are fitted by $b=\dfrac{\Delta y_F}{\Delta y_F+\varkappa}$ \cite{VecherninIndp}. Note, that the fact,
that pt-pt correlations are also in agreement with such fit,  pointing to 
considerably high density of strings in Pb-Pb collisions (such asymptotic was obtained in \cite{VechKol1} for AA 
collisions and it fails in pp case \cite{KovalenkoPoS}).
\begin{figure}[h!]
\centering
\includegraphics[width=11cm,clip]{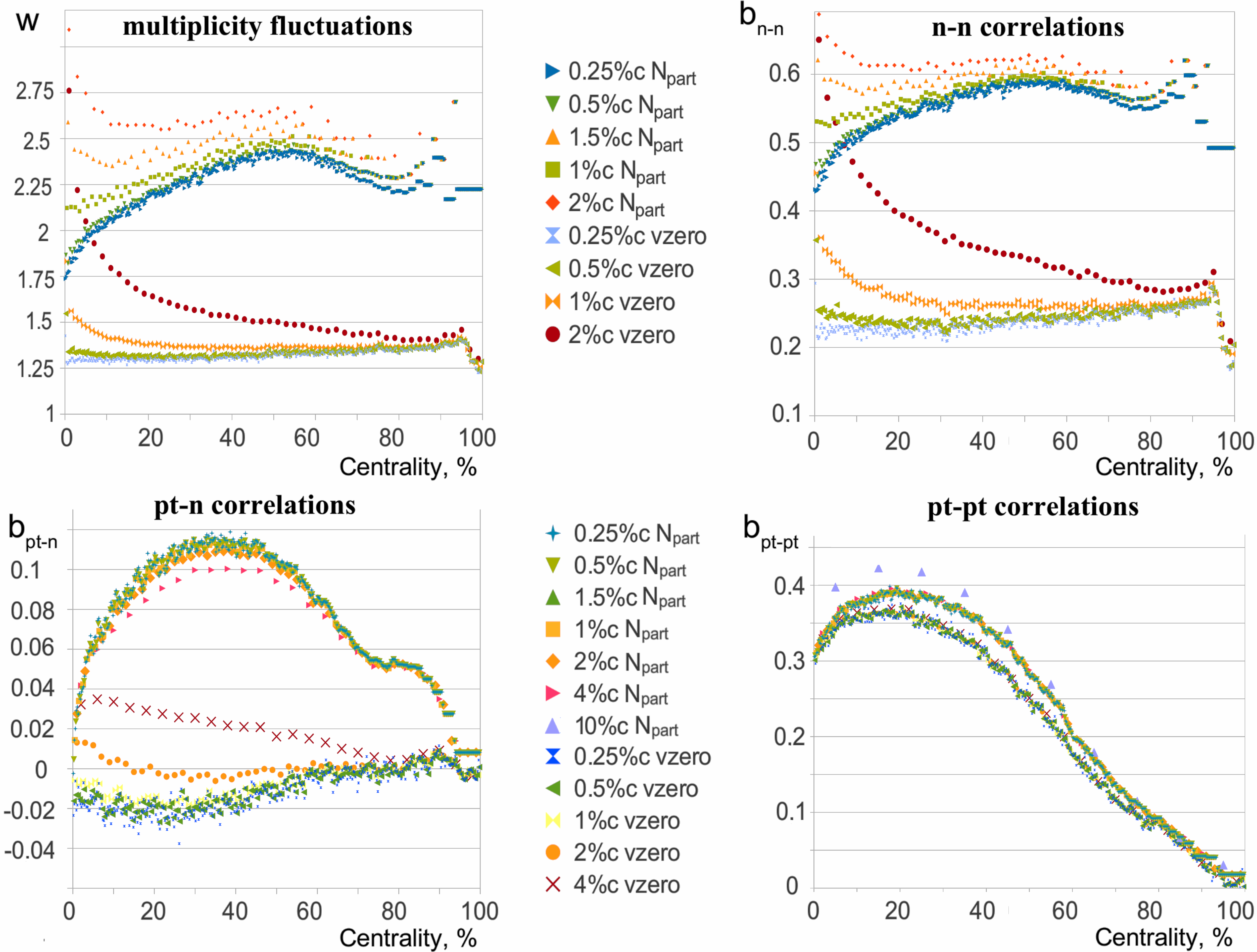}
\caption{Scaled variance of multiplicity $w$ in rapidity
window (0, 0.8), and $n$-$n$, $pt$-$n$, $pt$-$pt$ correlation coefficients in windows (-0.8,0),(0,0.8) as a function of centrality for different centrality class width  and methods of centrality  determination.}\label{fig-2}\end{figure}  

Results for multiplicity fluctuations $w$ and 
correlation coefficients as function of centrality
are shown in fig. \ref{fig-2}.
Both multiplicity fluctuations and $n\text{-}n$ correlations
are found decreasing with centrality for sufficiently
 narrow centrality bins, and in general their
 behaviour with centrality is similar in all cases.
 Important, that $w$ is higher for $N_{\text{part}}$-based centrality classes than for ``vzero''-based.
We note, that decrease of n-n correlations with centrality in model with string fusion contradicts to the predictions of CGC model  \cite{McLerannLRC}.

The results on pt-n correlations demonstrate always positive correlation coefficient for $N_{\text{part}}$-based centrality classes, what can be related with the fact that if at fixed number of participants many strings are produced, they undergo fusion and give higher pt, while at fixed
multiplicity in ``vzero'' windows (that means
almost fixed number of strings) $b_{pt\text{-}n}<0$ for narrow vzero-based centrality classes: the fixed multiplicity is achieved either from a few independent strings
 (pt lower) or from many-strings cluster (pt higher).
\begin{figure}[htb]
\centering
\tiny{\ \newline
\ \newline}
\includegraphics[width=6.0cm,clip]{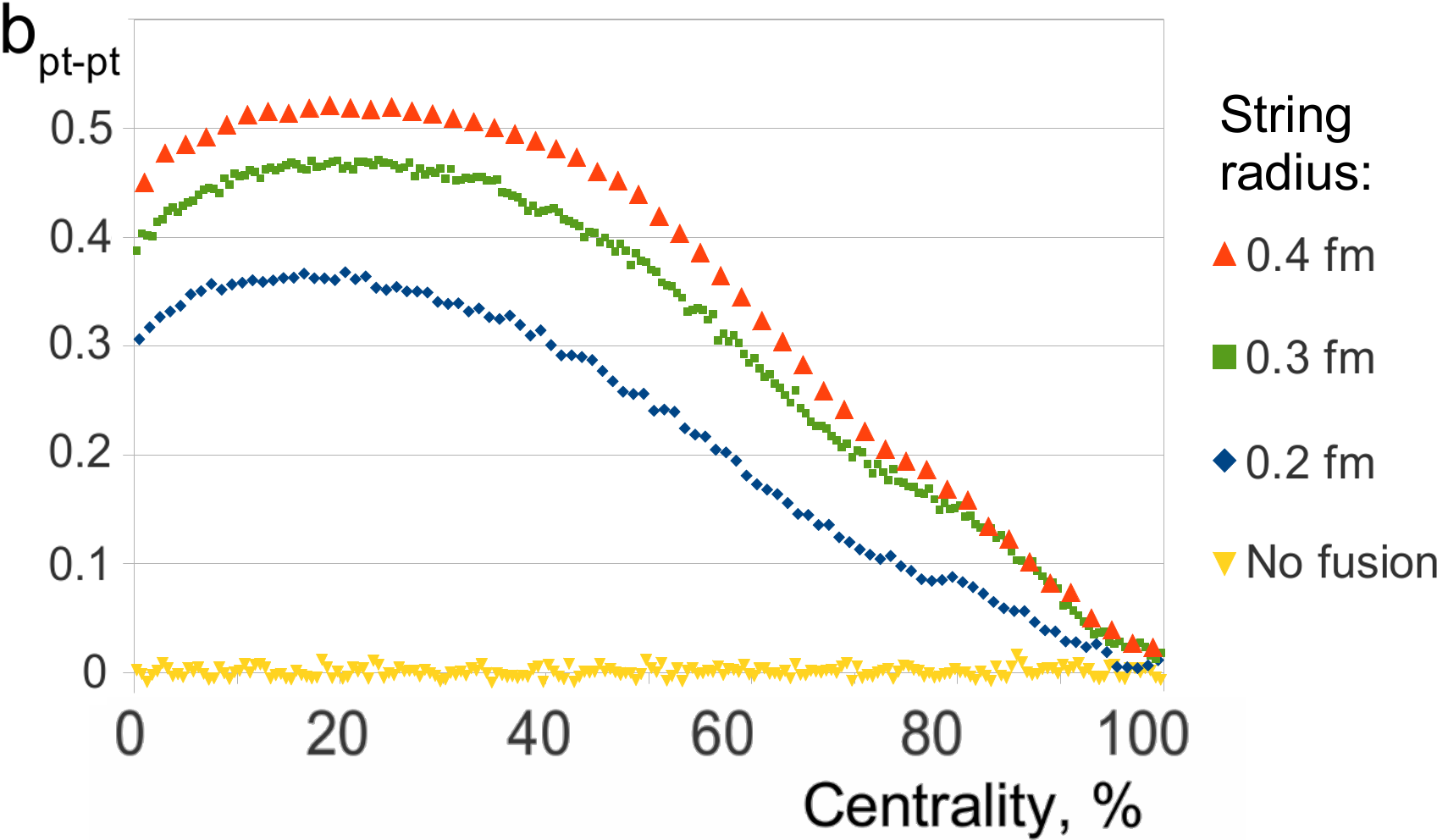}
\caption{Correlation between mean transverse momentum
of charged particles in rapidity window
(-0.8, 0) and (0, 0.8) for several values of string radius
from 0.2 fm to 0.4 fm (with string fusion) 
and also the case without string fusion.}
\label{fig-3}       
\end{figure}

The pt-pt correlations are found dominating over pt-n
and to have almost no effect from centrality determination issues.
The results are consistent in centrality bins narrower than 5\%. Thus this results could be easier compared with experimental data. At fig. \ref{fig-3} $b_{pt\text{-}pt}$ is 
plotted for different transverse radius of string. Pt-pt correlations are sensitive to this radius, and
the experimental data on $b_{pt\text{-}pt}$ correlations
would constrain its value.

 \begin{figure}[H]\centering
 \tiny{\ \newline
\ \newline}
\includegraphics[width=8.0cm,clip]{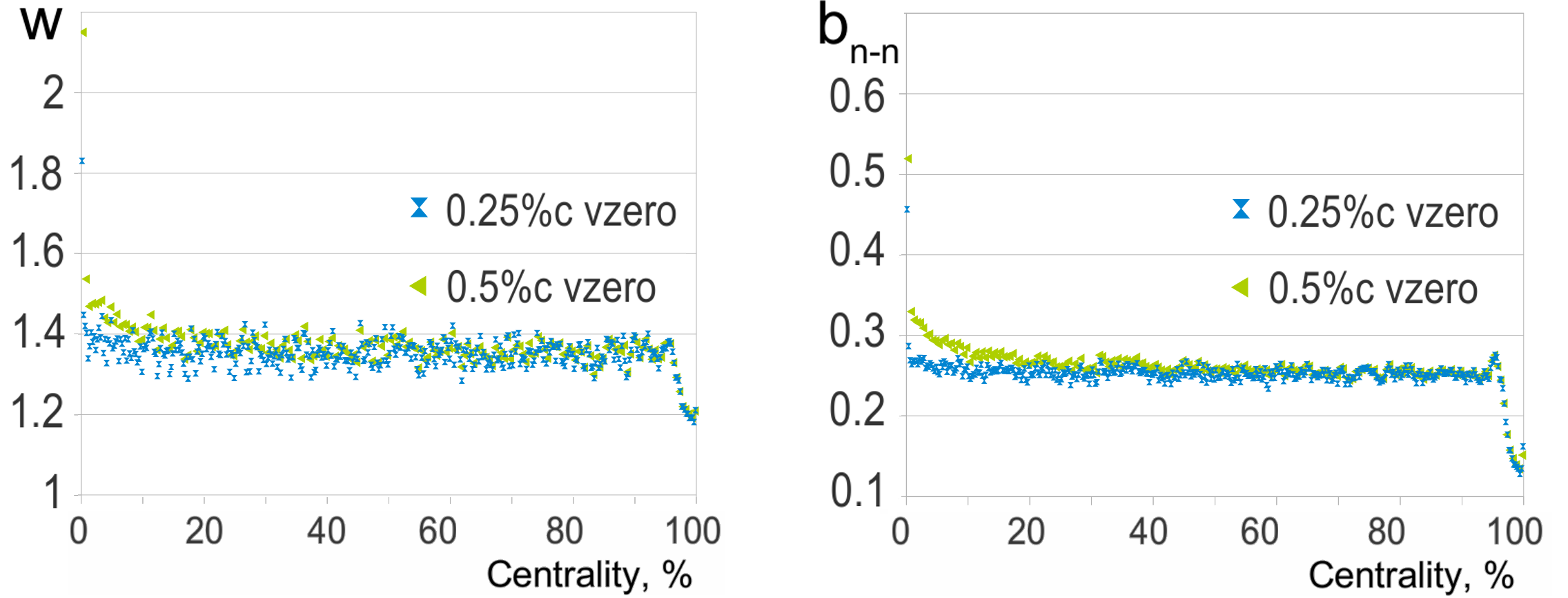}
\caption{Scaled variance of multiplicity $w$ (left) and correlation coefficient $b_{n\text{-}n}$ (right)  without string fusion. }
\label{fig-4}       
\end{figure}

Multiplicity fluctuations and $n\text{-}n$ correlations
are shown in fig. 4 for the case of no string fusion.
In narrow centrality bins they are constant with
centrality, that indicates that the decrease
of $w$ and $b_{n\text{-}n}$ is an effect of string fusion.
\section{Summary and conclusions}\label{concl}
Correlation coefficients and fluctuations strongly depend on the centrality bins and the methods of centrality determination. $b_{n\text{-}n}$ decreases with centrality for sufficiently small bin width in the model with string fusion, contrary to CGC model \cite{McLerannLRC}, 
and $w$ demonstrate very similar behaviour.
Pt-n correlations are negative for ``vzero'' and positive for $N_\text{part}$ centrality classes.
Pt-pt correlations have almost no effect from centrality determination issues and this is strongest prediction of string fusion model for experiment allowing to constrain the transverse radius of string.

\end{document}